\documentclass[12pt]{article} 

\usepackage{amsfonts}
\usepackage{graphicx}
\usepackage{dcolumn}
\usepackage{bm}
\usepackage{amsmath}

\renewcommand{\phi}{{\varphi}}

\newcommand{\vett}[1]{\mathbf{#1}} 
\newcommand{\interi}{\mathbb{Z}}

\title{Approach to equilibrium via Tsallis distributions 
  in  a realistic ionic--crystal model
	and   in the  FPU model}

\author{Andrea Carati\thanks{Dep. Mathematics, Universit\`a degli Studi di  Milano, 
                             Via Saldini 50, 20133 Milano -- Italy.}
        \and
        Luigi Galgani\thanks{Dep. Mathematics, Universit\`a degli Studi di  Milano, 
                             Via Saldini 50, 20133 Milano -- Italy.}
        \and   
        Fabrizio Gangemi\thanks{DMMT, Universit\`a di Brescia, Viale
                                Europa 11, 25123 Brescia -- Italy.}
        \and
        Roberto Gangemi\thanks{DMMT, Universit\`a di Brescia, Viale Europa 11, 25123 Brescia -- Italy.}
}
\date{}

\begin{document}

\maketitle
\abstract{
In Statistical Mechanics, Tsallis distributions were apparently
conceived in connection with systems presenting long--range
interactions. In fact, they were observed in numerical computations
for models of such a type, as occurring in the approach to equilibrium,
i.e., to a Maxwell--Boltzmann distribution.  Here we exhibit two
apparently new results.  The first one is that Tsallis distributions
occur also in an ionic--crystal model with long--range Coulomb forces,
which is so realistic as to reproduce in an impressively good way the
experimental infrared spectra.  Thus such distributions may be
expected to be actual physical features of crystals. The second result
is that Tsallis distributions occur in the standard short--range FPU
model too, so that the presence of long--range interactions is not a
necessary condition for Tsallis distributions to occur.  In fact, this is in
agreement with a previous result of the first author in connection with
the statistics of return times for the classical FPU model.  We thus confirm
the thesis advanced by Tsallis himself, that the relevant property
for a dynamical  system to present Tsallis distributions is that its dynamics
should be not fully chaotic, a property which is known to actually
pertain to long--range systems.
}
%

%
\maketitle

\section{Introduction}\label{uno}
In Statistical Mechanics, by Tsallis probability density (often
referred to as \emph {distribution}) one denotes a two-parameter
family of densities, which includes as a limit case the one--parameter
Maxwell-Boltzmann family. Denoting by $E$ the random variable under
consideration (for us, indeed, energy), the Tsallis family with
parameters $q \neq 1$ and $\beta>0$ has the form (with a normalization
factor $C$)
\begin{equation} 
f_{q, \beta}(E) = C \, \big[1+(q-1)\, \beta E \big]^{\frac 1 {1-q}}\ ,
\end{equation}
which reduces to the Maxwell--Boltzmann distribution $C\, e^{-\beta
  E}$ in the limit $q\to 1$.  Distributions of such a type were
apparently conceived as being suited for the statistical mechanics of
systems presenting long--range interactions.  In fact, Tsallis
distributions were actually observed in numerical simulations for
models of FPU type presenting long--range interactions involving all
particles, which should emulate Coulomb or gravitational interactions
\cite{tsallis1,tsallis2,tsallis3,rapisarda,ruffo,giansa}.
Starting from very peculiar non--equilibrium states, a transient state
was observed, pretty well described by a Tsallis distribution with
time--dependent parameters $q=q(t)$, $\beta=\beta(t)$, which for long
enough times converges to $q(t)=1$, i.e., to a Maxwell--Boltzmann
distribution, namely, to equilibrium. 

In the present paper we illustrate two apparently new results
concerning Tsallis distribution for systems of FPU type.  At variance
with the mentioned works, our results concern distributions of the
normal--mode energies rather than distributions of the particle
energies (or momenta).  Curiously enough it seems that, apparently,
investigations for distributions of normal--mode energies in FPU--type
systems were never performed in the more than seventy years elapsed
since the original FPU work.

The first result is that Tsallis distributions converging to a
Maxwell-Boltzmann one are met also for the normal--mode energies of a
realistic 3--$d$ FPU--like model. We are referring to an ionic crystal
model (actually, a LiF model) with Coulomb long--range interactions
(see \cite{litio1,litio2,litio3}), which has such a realistic
character as to reproduce in an impressively good way (and indeed
within a classical frame) the experimental infrared spectra. An
agreement between experimental data and theory over 9 orders of
magnitude for the infrared spectra of LiF, is exhibited in the first
two figures of the paper \cite{litio2}.  So the present result seems
to indicate that Tsallis distributions may be actual physical features
of crystals.

The second result originates within a more general frame, namely, the
dynamical foundations of statistical mechanics, investigated in terms
of the statistics of return times, with special attention to systems
which are not fully chaotic \cite{andrea1}. In such a frame, already
ten years ago it was pointed out by the first author \cite{andrea2}
that, in the classical nearest--neighbor FPU model, the statistics of
the return times is compatible with a Tsallis--type distribution in
the full phase space. In particular such a phase--space distribution
implies that the distribution of the normal--mode energies too be of
Tsallis type (albeit with different parameters). With such a result
for the return times in mind, we thus decided to investigate
numerically the approach to equilibrium of the normal--mode energies
in a classical FPU experiment (i.e., for a classical FPU model, and
for initial data with only a few low--frequency modes excited). The
result we found is that in such a case too the approach to equilibrium
occurs through a Tsallis distribution with time--dependent parameters,
albeit with some peculiarities with respect to the realistic
long--range model.

The common origin for the similar results (Tsallis distributions) in
the two different cases (long--range or short--range interactions) can
be caught at a dynamical level: namely, that in both cases one is
dealing with not fully chaotic systems.  The thesis is thus that
Tsallis distributions show up in dynamical systems which are not fully
chaotic. In particular such a class is well known to contain
long--range systems in the thermodynamic limit. So the present work
seems to support the thesis advanced by Tsallis himself, namely, that
''\emph{
  every time we have  a dynamics which is only weakly chaotic (typically
  at the frontier between regular motions and strong chaos), the need
  systematically emerges}`` for a $q$--statistics (see \cite{tsallislibro},
page 151). In other words,  the
original idea that Tsallis distributions show up in long--range
systems is correct, but it is the much more general property of
presenting only partially chaotic motions, that plays the relevant
role.

The new numerical results are illustrated in the next section
\ref{due}, and some concluding remarks then follow.

\section{The results} \label{due}
 
\subsection{The models}
\begin{figure}  
  ~\vskip -1.truecm
  \hspace{-0.6truecm}
  \includegraphics[width = 0.52\textwidth]{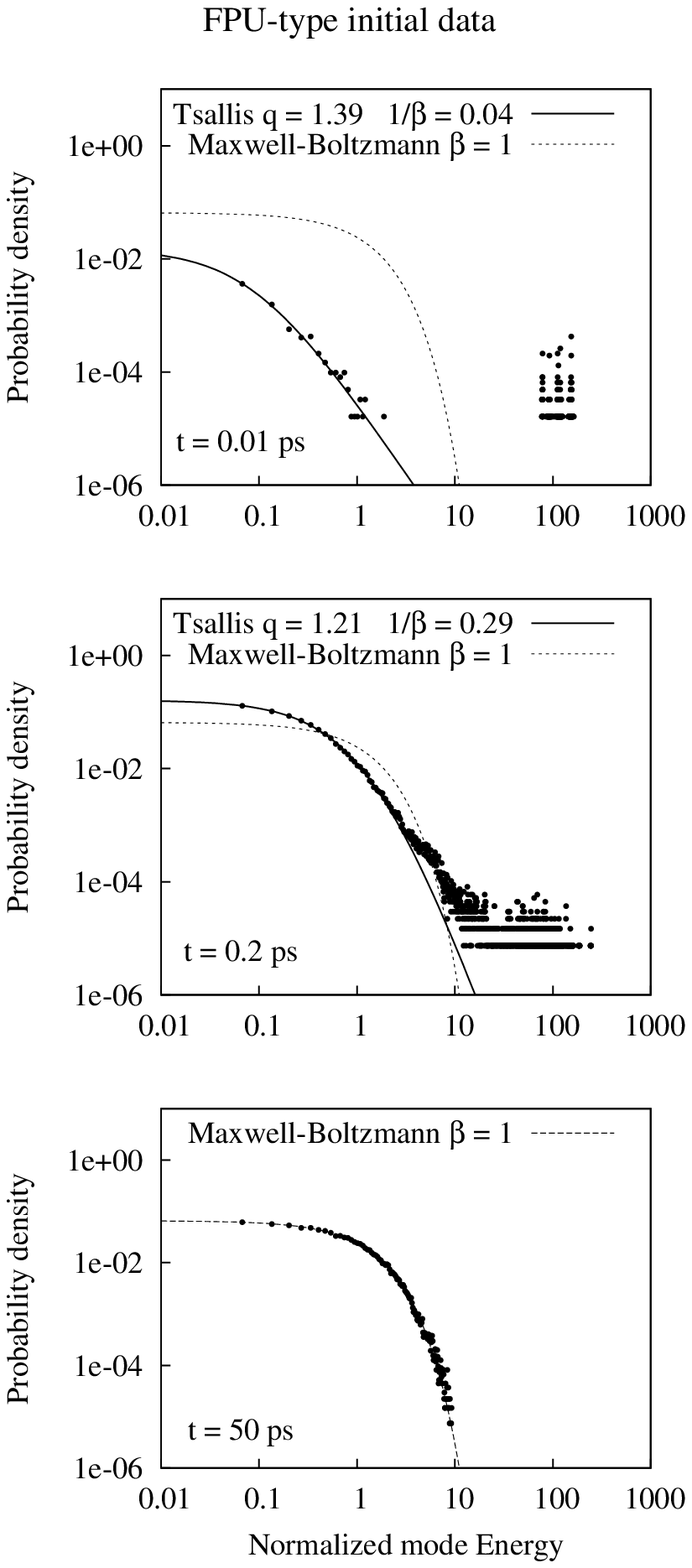}
  \hspace{0.1truecm}
  \includegraphics[width = 0.5\textwidth]{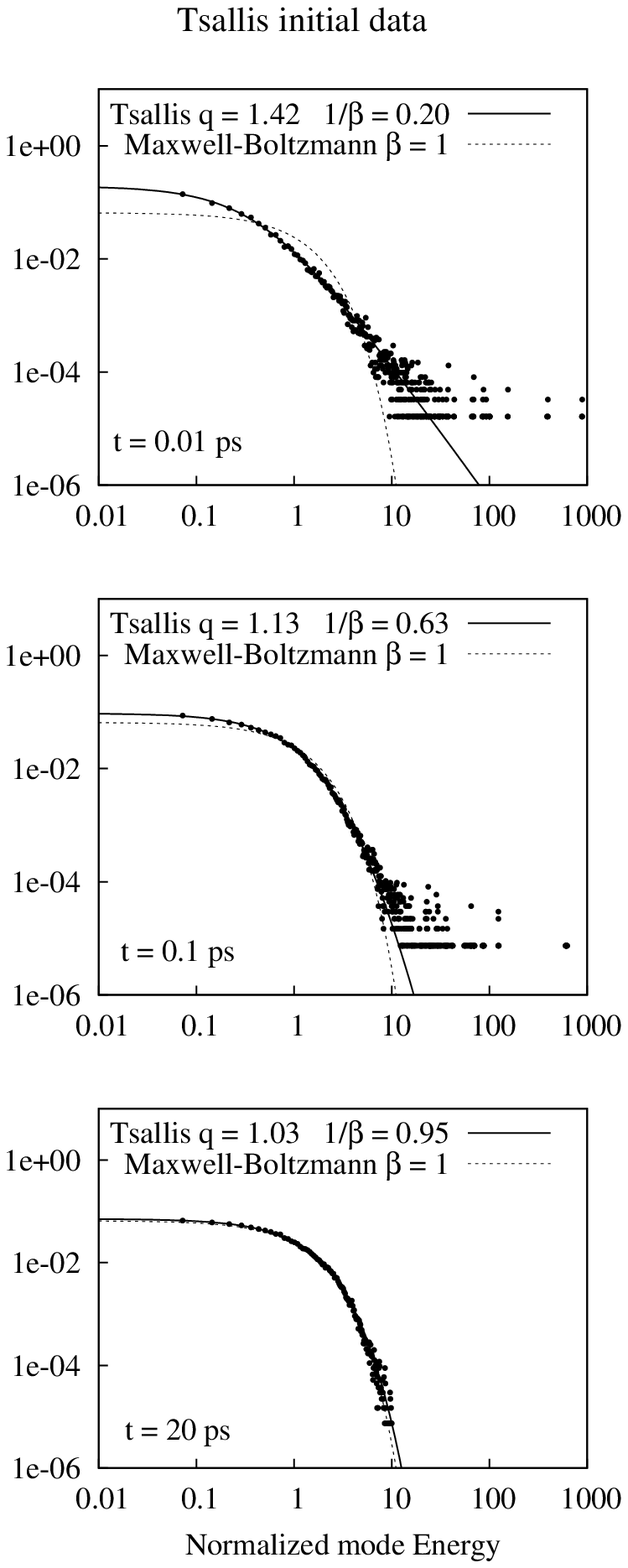}
  \caption{Histograms for the normalized normal--mode energies
   (i.e., their energies divided by the specific energy) at
    three different times (from top to bottom) for the realistic ionic
    crystal model: initial condition of FPU type (left) and of Tsallis
    type (right). Fit with a normalized Tsallis density (solid line),
    and comparison with a normalized Maxwell--Boltzmann density (dashed line).
    Number of particles $N=4096$.
    Specific energy $\varepsilon= 537$ K (left) and $\varepsilon= 501$ K (right).}
 \label{fig:1}
\end{figure}

We preliminarily add a few words about the models. The FPU model is
just the standard $\alpha-\beta$ one, which is universally known, and
is investigated here for $\alpha=1$ and $\beta=1$. We take
fixed end condition, so that the Hamiltonian reads
$$
 H = \sum_{i=1}^N \frac {p_i^2}2 + \sum_{i=0}^N V(x_{i+1}-x_i) \ ,
 \quad x_0=x_{N+1}=0 \ ,
$$
with $V(r)=\frac 12 r^2 + \frac {\alpha}3 r^3 + \frac {\beta}4 r^4$.

The realistic ionic crystal model is the standard one of Solid State
Physics, which was introduced long ago by the Born school.  In the Born
model (see for example \cite{bornhuang}) one considers $N$ ions in a
working cell of side $L$ with periodic boundary conditions, and one deals with the
ions as if they were point particles, interacting through pure Coulomb
forces (cared, as usual, through standard Ewald summations). The
contribution of the electrons, which don't show up in the model but
are known to produce polarization forces on the ions, is taken into
account in a phenomenological way by introducing a short--range
potential $V^{phen}$ acting among the ions, with suitable ``effective'' charges
substituted for the real ones. In our first paper \cite{litio1} the
phenomenological potential was just that originally proposed by Born,
namely, $V^{phen}(r)= C/r^6$, whereas a more complex potential,
depending on the pair of ions, was used in the subsequent papers
\cite{litio2,litio3}. In the end, the Hamiltonian reads 
$$
    H = \sum_{i=1}^N \frac {p_i^2}{2m_i} + \sum_{\vett k \in
      \interi^3} \sum_{i\ne j} V_{ij}(\vett x_{\vett k}^{ij}) \ ,
$$
where we have defined $x_{\vett k}^{ij} = |\vett x^i - \vett x^j -
  L\vett k|$, while $V_{ij}(\vett x)= e_ie_j/|\vett x| + V^{phen}_{ij}(\vett
  x)$.  Here $L$ is the side of the working cell, $\vett k$
   a vector with integer components.
More details can be found in the papers cited above.

\subsection{The results for the realistic ionic--crystal model}

We start with the results for the realistic model, which are
collected in Figure \ref{fig:1}, for two different initial conditions:
classical FPU type (left) and Tsallis type (right).
 The figures report the histograms of the normalized mode energies
(i.e., their energies divided by the specific energy)
at three different times, increasing from top to bottom. and
exhibit how a Maxwell--Boltzmann distribution is attained for
sufficiently long times.
In the left column the initial condition is of the classical FPU
type, with only a few low--frequency modes equally excited with random
phases, and vanishing energy to the remaining modes (actually,
all the modes having a frequency less than 100 cm$^{-1}$,
in number of 94 out of the total number 12288 of modes). 
In the right column, instead,
the initial condition is of Tsallis type (with $q = 1.4$ and $1/\beta
= 100$ K). In both cases one has $N=4096$, while the specific energy is
$k_BT$ with $T=537$ K (left) and $T=501$ K (right).

In the left column (FPU--type initial data) a rather impressive result is already exhibited in the
top panel. Indeed the panel corresponds to a time of just five
integration steps (each of 2 fs), and it shows that, at such a very
short time, the high energies remain
essentially gathered at the right side, whereas the small energies
constitute a separate group, and are already distributed pretty well according to a
Tsallis law. Such an immediate occurring of a Tsallis distribution
seems to be an interesting nonequilibrium phenomenon, which was
unknown to us. The central panel shows how at a subsequent time of 0.2
ps the two groups of energies (the small energies and the large ones)
start merging, so that the Tsallis fit, which is based just on
the low energies, now (at $q=1.21$) fails in the tail.  Eventually,
at time 50 ps, the histogram is very well fitted by the
Maxwell--Boltzmann distribution.  So, the small discrepancy of the
Tsallis fit at the intermediate time is just a peculiarity of the
particular nonequilibrium initial condition chosen.

It is thus quite natural to ask what occurs when the
initial energies themselves are  extracted according to a 
Tsallis distribution. This is exhibited in the right column, which
shows  that in such a case there exists a  time--invariance property, because the
distribution actually evolves within the Tsallis two--parameter family.
Notice that a very good fit with a Maxwell--Boltzmann distribution is
already attained at $t=20$ ps, which is at variance with the case of initial conditions of FPU type (left),
that requires a longer time (50 ps). The reason is that the Tsallis initial conditions
do not involve two different groups of energies.
An evolution within the Tsallis family is in agreement  with  results available in the literature,
and  seems to be  here   exhibited   in a  particularly neat way. 
\begin{figure}  
  \begin{center}
    ~\vskip -2.5truecm
    \includegraphics[width = 0.6\textwidth]{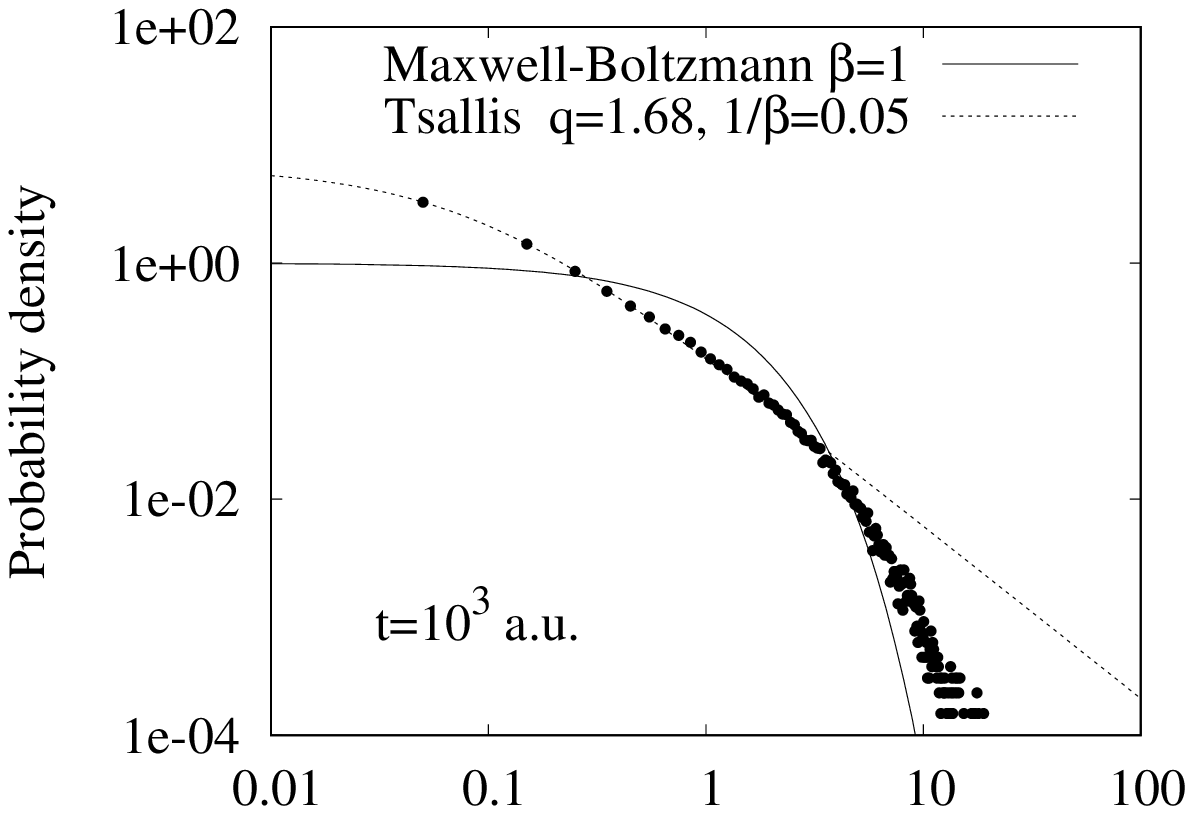}\\
    \includegraphics[width = 0.6\textwidth]{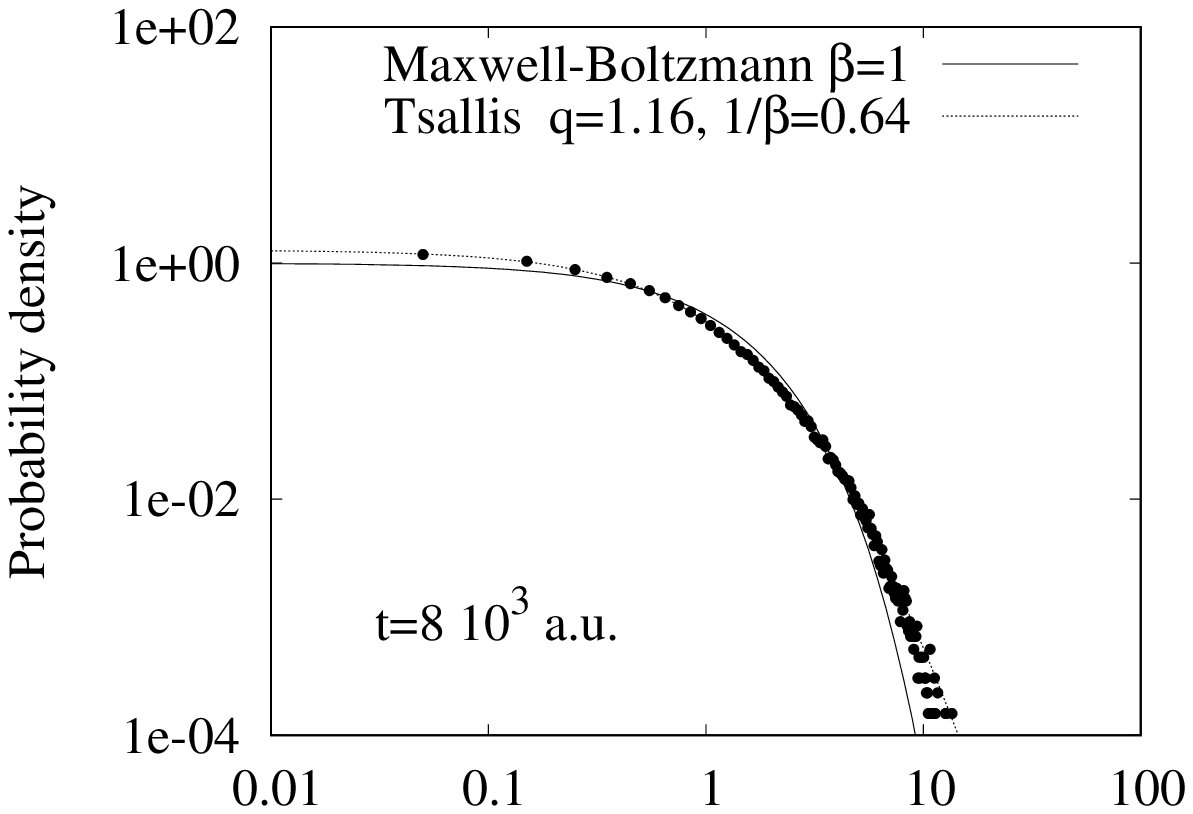}\\
    \includegraphics[width = 0.6\textwidth]{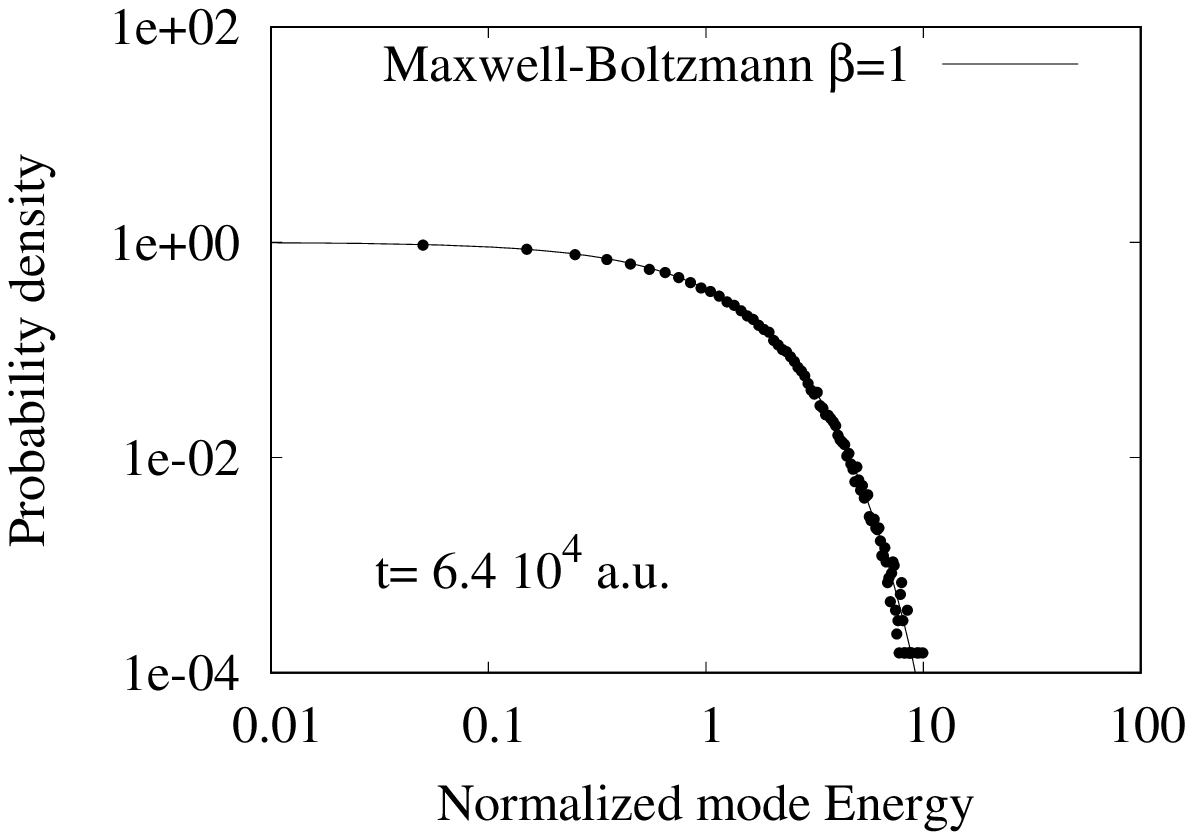}
  \end{center}\
  \caption{Same as Figure \ref{fig:1} for the classical nearest--neighbor
    FPU model. Initial condition with a few low--frequency modes excited.
The histogram refers only to the energies of the modes that were not initially excited.
The results are similar to those of the long--range realistic model,
apart from some details, discussed in the text.
 Number of particles $N=32768$. Number of initially excited modes: 512.
 Specific energy $\varepsilon= 0.0316 $.}
  \label{fig:2}
\end{figure}

\subsection{The results for the classical nearest--neighbor FPU model}

We report now, summarized in Figure \ref{fig:2}, the results  for the
classical nearest--neighbor FPU model. We consider a system of
$N=32768$ particles with fixed ends, and we  initially excite a 
packet of low frequency modes, which contains the
512  lowest ones, with equal energies and random phases. The specific
energy was fixed at $\varepsilon= 0.0316$, a value for which the
equipartition time, according to the paper \cite{benettiponno}, is of
order of $5\cdot 10^2$. We will consider here larger times because, obviously,
the occurrence of a single distribution for the energies of any frequency,
implies equal mean energy for all frequencies, i.e., equipartition.

In the  figure we report, in the different panels corresponding to increasing
times, the distributions of the normalized energies of the modes which were not
initially excited. The upper panel refers to the distribution after
$1.25 \cdot 10^5$ integration steps (which, with our choice of the
integration step, amounts to $t\simeq 10^3$), the central panel refers to a
time of  $10^6$ integration steps ($t\simeq 8\cdot 10^3$), and the lowest one
 to $8\cdot 10^6$ integration steps ($t\simeq 6.4\cdot 10^4$). Each
histogram was obtained using the data of four different trajectories,
corresponding to 
different random choices of the phases of the excited modes.

The upper panel shows  that at a short time the energy
distribution is well fitted by a Tsallis one, apart from the high--energy
tail, namely apart from energies larger than 3 times the specific energy, which appear to be
exponentially distributed. As  time increases, the crossover energy
increases, and at a certain time (central panel) a Tsallis distribution
fits well the histogram  over the whole energy range. Such a
distribution eventually becomes a Maxwell--Boltzmann one, as shown in
the bottom panel. 

The results are thus essentially similar to those of the long--range
case, apart from the fact that the occurrence of a Tsallis distribution
requires a larger time scale. This is due to a fact 
that was observed  since the first works on the FPU model. 
Namely, that for standard FPU--type
initial conditions the dynamics builds up a rather stable low--frequency
packet, in which the energies of the modes 
decay exponentially for increasing  frequencies 
(see for example  Table I of the paper \cite{gs}), so that 
   the energy distributions too
present an exponential tail persisting for rather long times. This
fact seems to explain the crossover between the Tsallis distribution for
the low energies and the exponential tail of the high energies, 
a crossover  that shifts
towards the large energies as time increases. 
In conclusion, at variance
with the long--range case, at very short times  the Tsallis distribution
should occur here too, but only for a very small range of low energies.

Anyhow, long--range interactions
 seem not to  be necessary for the occurring of Tsallis  distributions
evolving towards a Maxwell--Boltzmann distribution. Moreover, another
interesting fact is that the attainment of equipartition does not guarantee
the attainment of equilibrium (see also \cite{danieli,danieli2}).

\section{Conclusions}\label{ultimo}
The main result of the present paper seems to be  that Tsallis distributions 
are observed during  the approach to 
equilibrium for the normal--mode energies of FPU--like systems,
not only in the case of  long--range interactions for which they were 
apparently conceived,
 but also in the case of short--range interactions, albeit with some
 peculiar features in the latter case.
 Such a result   might have been expected,  on the basis of 
 previous results for  the statistics of return times in dynamical
 systems,  obtained  in the   frame  of 
the dynamical foundations of  statistical mechanics. From such results
it is confirmed  that the  relevant dynamical feature for the occurring of  
Tsallis distributions should be   lack of full chaoticity, which is
the thesis advanced by Tsallis himself, for example  in \cite{tsallislibro}.

A final comment concerns a metastability phenomenon for which indications
were given in previous works in the long--range case (see for example
ref. \cite{tsallis2}). Namely, the function $q=q(t)$ would present a plateau,
and moreover the length of the plateau would strongly increase for increasing $N$.
This is a very interesting perspective, but the results available to us in the realistic
model are not yet sufficiently clear in this connection.

\section*{Acknowledgements}
Funding: research carried out with the support of resources of
Big\&Open Data Innovation Laboratory (BODaI-Lab), 
University of Brescia, granted by Fondazione Cariplo and Regione Lombardia.

\end{document}